\documentclass[aps,prb,twocolumn,amsmath,amssymb]{revtex4}
\usepackage{graphicx,times}
\usepackage{bm}
\usepackage{color}
\usepackage{multirow}
\usepackage{times}

\begin{document}
\title{
Impurity-doped Kagome Antiferromagnet~:~A Quantum Dimer Model Approach
}
\author{Didier Poilblanc$^{1}$ and
Arnaud Ralko${^2}$ }
\affiliation{
${^1}$ Laboratoire de Physique Th\'eorique UMR5152, CNRS and Universit\'e de Toulouse, F-31062 France \\
${^2}$ Institut N\'eel UPR2940, CNRS 
and Universit\'e de Grenoble, F-38000 France
} 
\date{\today}
\begin{abstract}
The doping of quantum Heisenberg antiferromagnets on the kagome lattice by non-magnetic impurities is
investigated within the framework of a generalized quantum dimer model (QDM) describing a) the valence bond crystal (VBC),  
b) the dimer liquid and c) the critical region on equal footing.
Following the approach by Ralko et al.
[Phys. Rev. Lett. {\bf 101}, 117204 (2008)] for the square and triangular lattices, we introduce the (minimal) extension of the
QDM on the Kagome lattice
to account for spontaneous creation of {\it mobile} S=1/2 spinons
at finite magnetic field. Modulations of the dimer density (at zero or finite magnetic field) and of the local field-induced magnetization in the vicinity of 
impurities are computed using Lanczos Exact Diagonalization techniques on small clusters ($48$ and $75$ sites).  The VBC is clearly revealed from its pinning by impurities,
while, in the dimer liquid, crystallization around impurities involves only two neighboring dimers.
We also find that a next-nearest-neighbor {\it ferromagnetic} coupling favors VBC order. 
Unexpectedly, a small size spinon-impurity bound state appears in some region of the the dimer liquid phase.
In contrast, in the VBC phase the spinon delocalizes within a large region around the impurity, revealing the weakness of the VBC confining potential.
Lastly, we observe that  an impurity concentration as small as 4$\%$ enhances dimerization substantially.
These results are confronted to the Valence Bond Glass scenario [R.R.P.~Singh, Phys. Rev. Lett. {\bf 104}, 177203 (2010)]
and implications to the interpretation of the Nuclear Magnetic Resonance spectra of the Herbertsmithite 
compound are outlined.

\end{abstract}
\maketitle


\section{Introduction}

\begin{figure}
\includegraphics[width=0.90\columnwidth,clip]{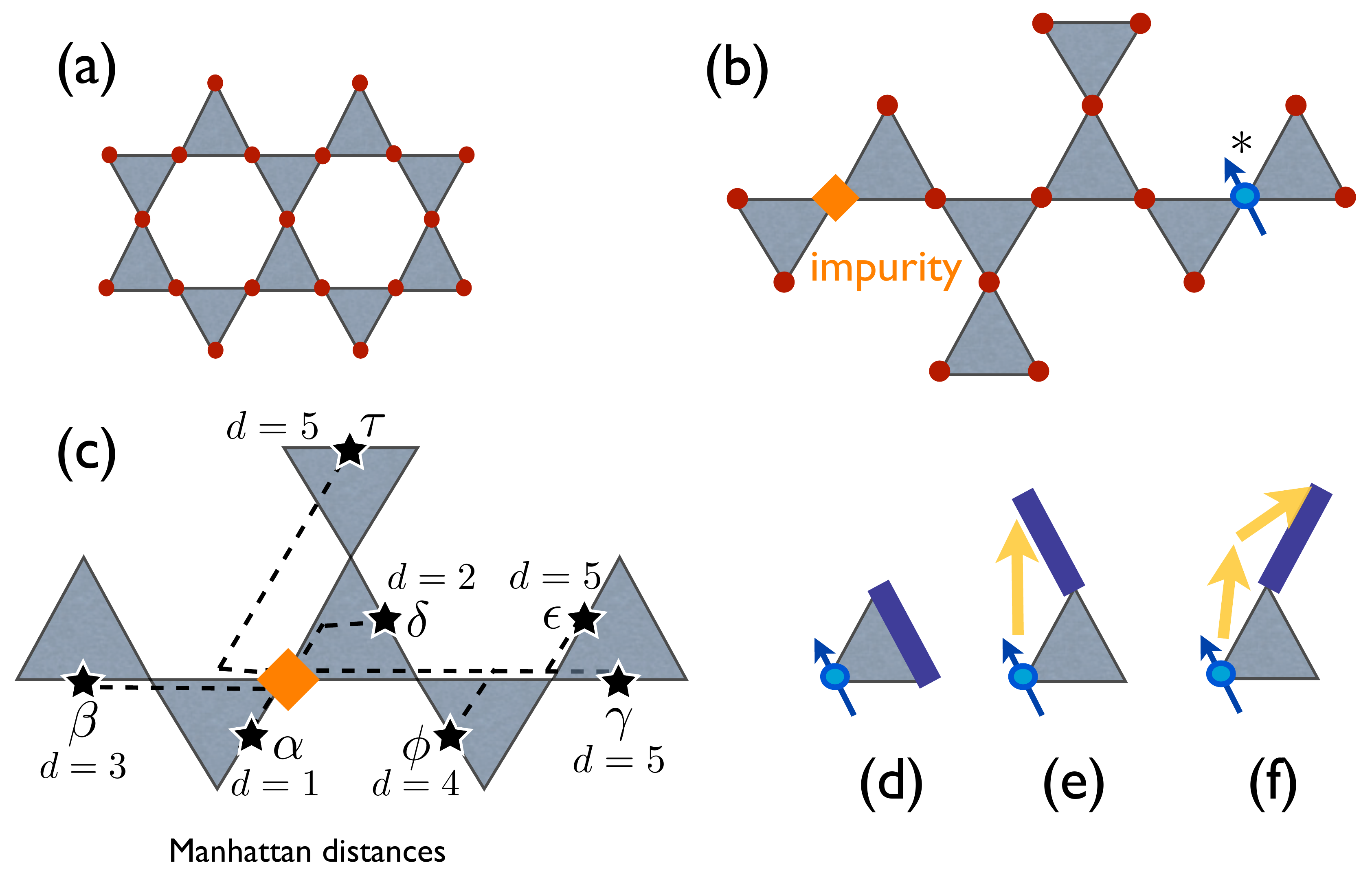}
\caption{\label{Fig:intro} (Color online). 
(a) Kagome lattice consisting of a hexagonal lattice of corner-sharing frustrated triangular units. 
The Heisenberg AFM interaction acts on all bonds of every triangle.
(b) A non-magnetic static impurity (lozenge) in the kagome lattice is described as a vacancy.  Mobile spinons (arrows) carry spin-1/2 degrees of freedom.
(*) The $N=75$ cluster contains only a single impurity and no spinon.
(c) The Mahattan distance $d$ is obtained by counting the number of unit segments (oriented along the two crystallographic axis)
on the path joining the impurity to a given bond center. Non-equivalent bonds are labelled by greek letters. 
(d)  A spinon (carrying spin-1/2 shown by a blue arrow) on a triangle cannot hop on the sites of the same triangle.
(e-f) A spinon can hop at  further distances leading to a 120-degree (c) or 180-degree (back-flow) rotation (d) of the involved dimer.}
\end{figure}

The spin-1/2 quantum Heisenberg antiferromagnet (QHAF) on the kagome lattice
(see Fig.~\ref{Fig:intro}(a)) is believed to be the paradigm of frustrated
quantum magnetism. It is also the best candidate to observe exotic
(non-magnetically ordered) phases predicted by theorists.  Among those, an algebraic spin
liquid~\cite{algebraic}, a gapped dimer --or spin-- liquid~\cite{leung,dmrg} or a number of valence bond crystals (VBC) with
12-site~\cite{maleyev}, 18-site~\cite{marston}  and 36-site~\cite{marston,senthil,singh} unit cells or with columnar dimer 
order~\cite{auerbach} have all been proposed as possible ground state (GS) of the kagome QHAF. 

On the experimental side, 
Herbertsmithite~\cite{herbertsmithite} is an almost perfect
realization of a kagome QHAF.  In this material Copper atoms
carrying spin-1/2 degrees of freedom (oxidation state $+2$) 
are located on the
sites of (very weakly coupled)  kagome layers.  In addition to the
crystallographic structure, the closeness to an ideal system  is
supported by the remarkable {\it absence} of any magnetic ordering at the
lowest attainable temperatures~\cite{no-order}, as expected for a highly frustrated quantum
magnet. 
 
 However, small deviations from a perfect kagome QHAF are known in 
 this material.  In addition to a small Dzyaloshinskii-Moriya (DM) anisotropy~\cite{dm}, 
 Herbertsmithite contains a small amount of
 non-magnetic Zinc impurities substituting Copper atoms in the kagome planes.
 Nuclear Magnetic Resonance (NMR) offers a fantastic tool to probe impurities
 via the change induced by a nearby impurity on the local magnetic field at the
 nucleus whose resonance is being measured.  The first estimations from NMR~\cite{nmr} of
 the Zinc concentration range from 6 to 10 $\%$.  It is not clear therefore,
 which experimental features in Herbertsmithite can really be attributed to
 intrinsic properties of the kagome QHAF and which of them are induced e.g. by Zinc
 substitution.

One of the strong motivations of this work is to investigate the effect of spinless impurities
in quantum disordered (i.e. non-magnetic) phases which are the most serious candidates
for the Herbersmithite ground state (GS). 
Starting from a simple description of this material in terms of a
SU(2)-symmetric kagome spin-1/2 QHAF (neglecting the small DM
anisotropy) recent calculations of NMR spectra~\cite{rozenberg} seem to reflect key experimental features~\cite{nmr}.
On the other hand, the local susceptibilities (Knight shifts) obtained by high temperature series expansion~\cite{motrunich} were claimed to agree with 
an algebraic (gapless) spin liquid phase. Here we provide an alternative analysis based on a generalized quantum dimer model (QDM) which can naturally describe  a) topological  
(so-called $\mathbb{Z}_2$) spin-liquid, b) VBC and c) the critical region 
on equal footing and distinguish features around impurities which are specific to each phase.
The paper is organized as follow: first, in Section~\ref{Sec:model}, we describe the QDM framework, the modeling of the 
static dopants and the simplest extension of the QDM to account for a small magnetization (induced by a finite magnetic field) in the ground state.
In Section~\ref{Sec:single}, the results for a single imbedded impurity are described. This includes the investigation of the pinning of the VBC and the calculation of the
spin density of a doped mobile spinon around the vacancy.
Then, in Section~\ref{Sec:finite}
we move on to the case of several impurities and show how the dimerization pattern can be enhanced. A discussion of the experimental NMR data in the light of our results
is given in Section~\ref{Sec:nmr} as well as some concluding remarks  in Section~\ref{Sec:conclusions}.

\section{Theoretical description} 
\label{Sec:model}

\subsection{Impurities described as vacancies}

Let us first discuss qualitatively the effect of a single 
impurity~\footnote{In the very diluted limit, the magnetic response scales with the impurity concentration.}. 
A Zinc impurity is a spinless ion and therefore can be described as a vacant site as depicted in Figs.~\ref{Fig:intro}(b,c). 
In other words, the bond exchange interaction ${\bf S}_i\cdot {\bf S}_j$ (in units of the exchange coupling $J$) acts on all 
bonds of the kagome lattice except on the four bonds connected to the impurity. 
Using Lanczos exact diagonalizations (LED) of small clusters of the kagome QHAF~\cite{pure}, it has been 
shown that a single impurity 
tends to `localize' two singlet bonds next to it~\cite{dommange,sindzingre,ioanis}.
Although this local phenomenon is well captured even on very small clusters~\cite{rozenberg}, LED of the QHAF do not allow 
to investigate reliably the perturbation of the media at larger distances from the impurity, which can be probed e.g. by NMR. 
Therefore, we shall use here an effective description based on the recently developed QDM
allowing to handle larger clusters\cite{ralko,schwandt}.

\subsection{Competing singlet phases in the pure system} 

\begin{figure}[htb]
\begin{center}
  \includegraphics[width=0.85\columnwidth,clip]{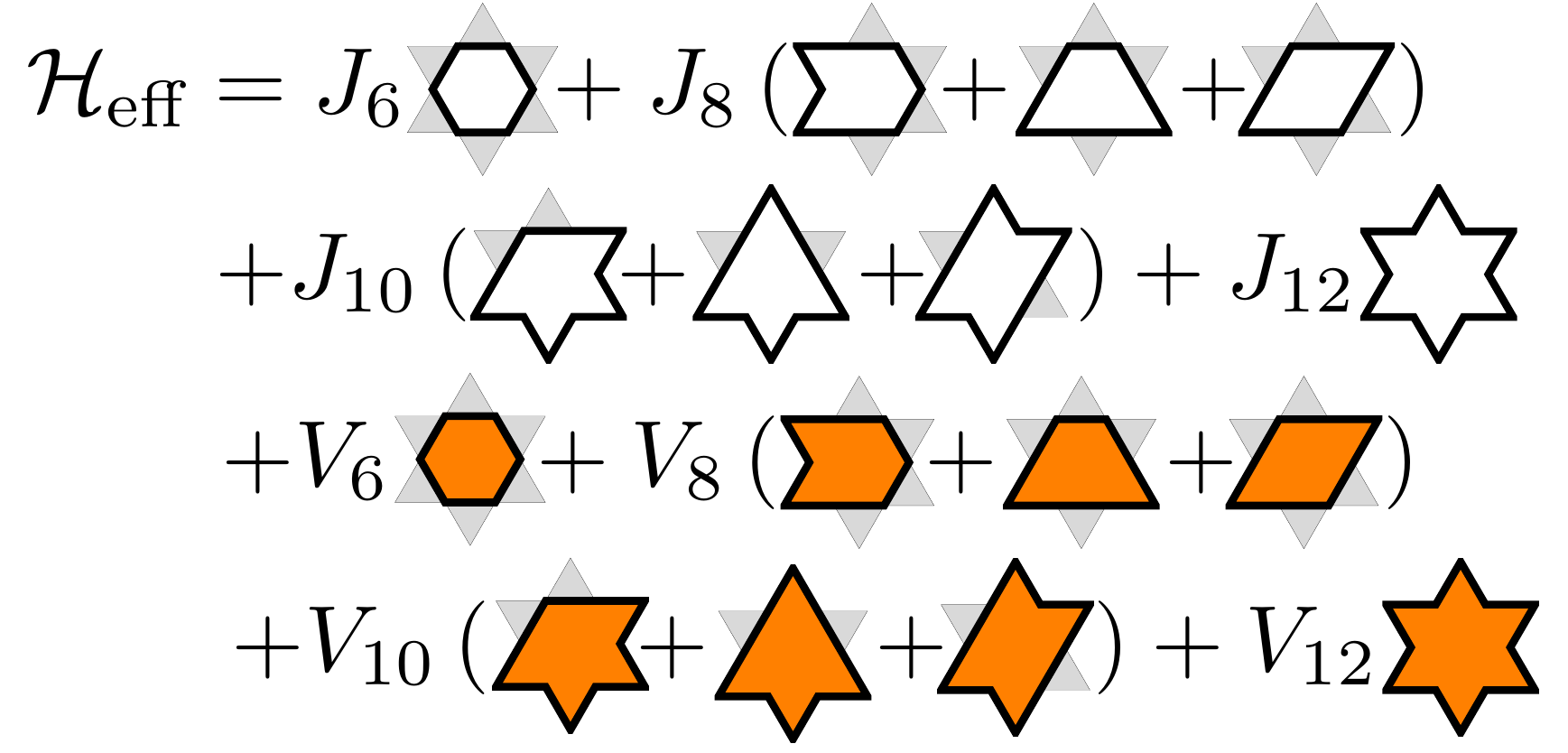}
\end{center}
\caption{The effective QDM used here. A sum over all the hexagons of the lattice is implicit.
Kinetic terms ($J_\gamma$) promote cyclic permutations of the dimers
around the loops (shown as thick lines) and diagonal 
terms ($V_\gamma$) count the numbers of ``flippable'' loops. }
\label{Eq:heff}
\end{figure}

Starting from a projection of the QHAF on the
(non-orthogonal) NN singlet basis~\cite{mambrini}
(an approximate resonating VB description)
followed by a transformation onto
an ad-hoc orthogonal `dimer' basis and, finally, by a series expansion (in terms of a small overlap parameter $\alpha=1/\sqrt{2}$), 
one gets
an effective QDM of the type shown in Fig.~\ref{Eq:heff}.
Here we use the amplitudes obtained in Ref.~\onlinecite{schwandt}, $V_6= 1/5$, $V_8=2/63$, $V_{10}=1/255$, $V_{12}=0$, $J_6=-4/5$,
$J_8=16/63$, $J_{10}=-16/255$, $J_{12}=0$ (in units of $J$) which defines our "Heisenberg" effective QDM 
$H_{\rm Heis}$. For convenience, the labels $\gamma$ in the amplitudes $J_\gamma$ and $V_\gamma$  correspond to the lengths of the 
associated resonance loops. An exactly solvable "Rokhsar-Kivelson" (RK) QDM $H_{\rm RK}$ defined by vanishing diagonal amplitudes $V_\gamma=0$ 
and equal $J_\gamma$ kinetic amplitudes (set here to -1/4) has also been introduced~\cite{Z2-RK}. Following Ref.~\onlinecite{poilblanc}, we shall consider
a linear interpolation between the two Hamiltonians, $H_{\rm eff}(\lambda)= \lambda H_{\rm Heis}+ (1-\lambda) H_{\rm RK}$. 
As we shall see next, it is remarkable that this "generalized" QDM  can describe topological  
dimer-liquid, VBC and the critical region on equal footing. Whether this model (for $\lambda=1$ or around this value) describes faithfully the original (unprojected) microscopic QHAF on the Kagome lattice
is still under debate. However, lots of insights on how impurities behave {\it generically} in the
quantum disordered phases described by this effective model can be obtained.

\begin{figure}[htb]
\begin{center}
  \includegraphics[width=0.95\columnwidth,clip]{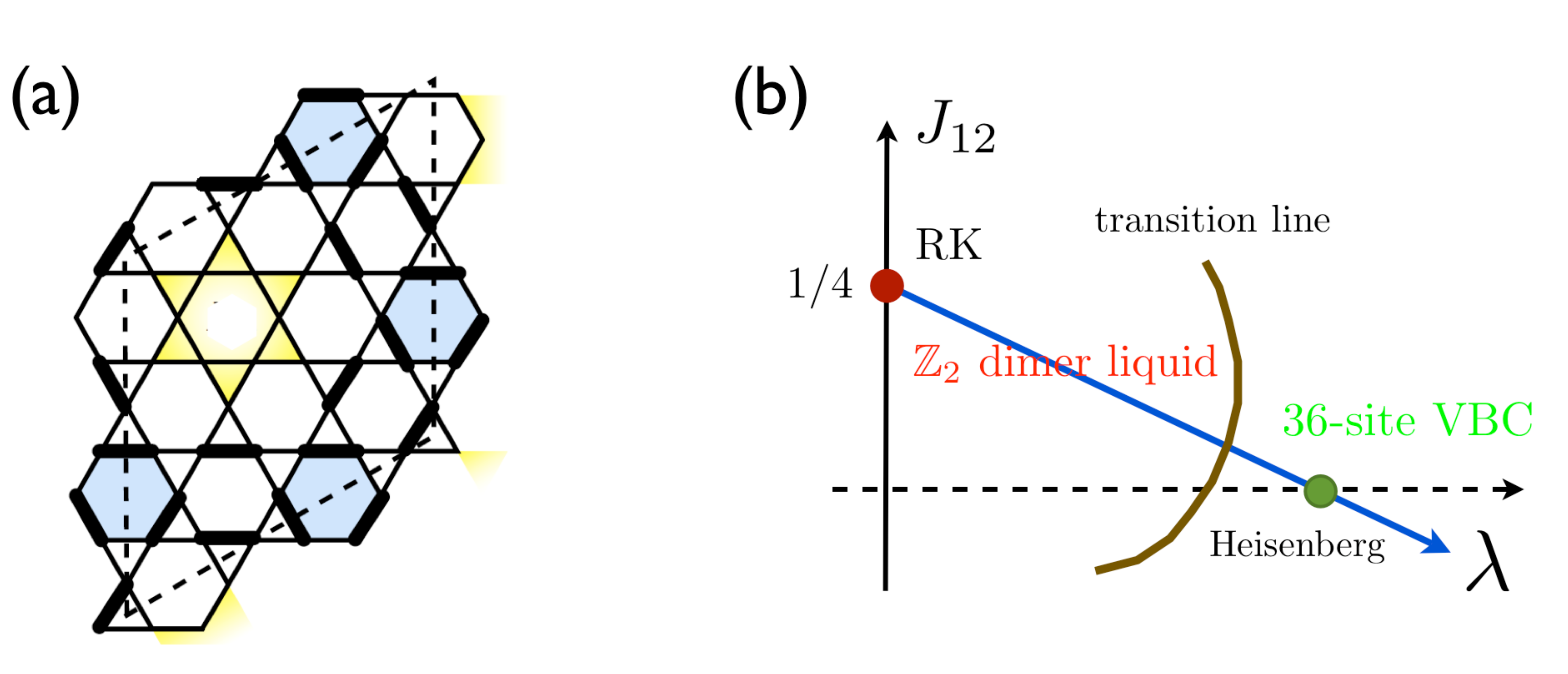}
\end{center}
\caption{(a) Schematic representation of the 36-site VBC supercell (from Refs.~\onlinecite{schwandt,poilblanc}). (b) Phase diagram of the {\it pure} QDM.
In the present study, one moves along the (blue) $\lambda$-axis.}
\label{Fig:phasediag}
\end{figure}

So, prior to the introduction of impurities, let us give a brief summary of the results of the pure  generalized QDM (in zero magnetic field).
Generically, short or long-range ordering of the dimers can happen yielding to dimer liquids or VBC, respectively.
Numerical results~\cite{poilblanc} show that a 36-site unit cell VBC (shown schematically in Fig.~\ref{Fig:phasediag}(a)) is stabilized for  $\lambda> \lambda_{\rm C}\simeq 0.94$, in particular at the
"Heisenberg" point $\lambda=1$.
A remarkable (topological) gapped $\mathbb{Z}_2$ dimer liquid which bears an exact analytical form at the RK point $\lambda=0$ is stable for $\lambda<\lambda_{\rm C}$
as shown in the phase diagram of Fig.~\ref{Fig:phasediag}(b).
Besides, at exactly $\lambda=1$, $J_{12}$ vanishes so that the undetermined 
chiralities of the pinwheel (or star) resonances of the VBC pattern (see e.g. Ref.~\onlinecite{poilblanc} for details)  lead to an extra (Ising-like) macroscopic degeneracy. 
These findings at $\lambda=1$ are in perfect agreement with recent series expansions on the QHAF~\cite{singh}.
Note that the proximity to a Quantum Critical Point (QCP) suggests that dimer fluctuations remain strong and that the VBC amplitude is weak at $\lambda=1$.
However it should be stressed here that the derivation of the $\lambda=1$ effective QDM is only semi-quantitative since it relies i) on the projection
of the Heisenberg Hamiltonian onto the singlet subspace spanned by the NN VB and ii) it is subject to a (controlled) truncation of a series expansion. 
Besides, the real material might contains longer range exchange interactions. This suggests that a faithful description of the material in terms of a generalized QDM might not correspond exactly to $\lambda=1$. We can however consider $\lambda$ as a "phenomenological" parameter which enables to "tune"
the system in one of the two quantum disordered phases or in the critical region in between to investigate the generic role of impurities
in each case.

\subsection{Extending the QDM to finite magnetic field}

{\it Spinon hoppings --}
In an NMR setup a small magnetic field is applied to slightly polarize the system.
The local site-dependent magnetization is then experimentally accessible via the measured Knight shift.
However, ground states with finite magnetization cannot be addressed theoretically within the above QDM which only 
describes the singlet subspace dynamics. 
Nevertheless, a finite field/magnetization setup can be realized theoretically by introducing extra spin-1/2 degrees of freedom (named "spinons" hereafter) in the QDM description 
as first proposed in Ref.~\onlinecite{spinons}. Physically, such a
spinon (polarized along the magnetic field) can be viewed as resulting from the breaking of a singlet bond by the introduction of 
a vacant site (impurity). Spinon nearest-neighbor pairs can also appear or disappear by exchanging
with dimers (see below). The bond exchange interaction of the QHAF leads to the motion of the spinons followed by some
dimer "backflow"~\cite{hao}. 
Since, the generalized QDM has been derived in Ref.~\onlinecite{schwandt} assuming a fermion representation of the (original) SU(2) dimers,
consistency implies that 
the mutual (bare) statistics of
the spinons should be fermionic~\footnote{Note the doped spinon (holon) QDM is invariant under simultaneously (i) reversing the signs of {\it all} kinetic 
amplitudes $J_\gamma$ and (ii) changing the
statistics of the spinons (holons) from fermionic (bosonic) to bosonic (fermionic). See D.~Poilblanc, Phys. Rev. Lett. {\bf 100}, 157206 (2008) for the square lattice.}.
Here, we extend the procedure of Ref.~\onlinecite{schwandt} to derive from the microscopic QHAF the simplest effective Hamiltonian (in lowest $\alpha^2$ order) governing the 
dynamics of the (fermionic) spinons sufficient to capture the main features of spinon delocalization. 
The relevant processes are shown in Figs.~\ref{Fig:intro}(d-f). The overlap matrix elements between the initial and final SU(2) configurations $|\phi\big>$ and $|\psi\big>$ involved in these
processes  are ${\cal O}_{\phi,\psi}=-1/2$. The corresponding Heisenberg matrix elements ${\cal H}_{\phi,\psi}$ equal +1/2 (in units of $J$) for the last two processes of Figs.~\ref{Fig:intro}(e,f)
and vanishes for the first one of Fig.~\ref{Fig:intro}(d). Note that, for consistency with the definition of the QDM of Fig.~\ref{Eq:heff}, we 
subtract a singlet-bond energy and include a 4/3 multiplicative factor~\cite{note-rescale}. 
Using the expansion scheme given by (the first two terms of) Eq.~(41) in Ref.~\onlinecite{schwandt}, we 
deduce easily that, in the effective QDM, the first process vanishes and the two remaining kinetic processes shown in Figs.~\ref{Fig:intro}(e,f) involve
the same "spinon hopping" amplitude $t_{\rm sp}=+1/2$. This procedure also generates (at order $\alpha^4$) a potential term of amplitude $V_{\rm sp}=1/4$ 
which "counts" the number of possible "spinon hopping" of each kind. Stricly speaking, this (minimal) effective spinon Hamiltonian is most relevant to the physical case $\lambda=1$
but it is also interesting to investigate its properties in an extended range e.g. $0.5\le \lambda\le 1$. 

\begin{figure}[htb]
\begin{center}
  \includegraphics[width=0.95\columnwidth,clip]{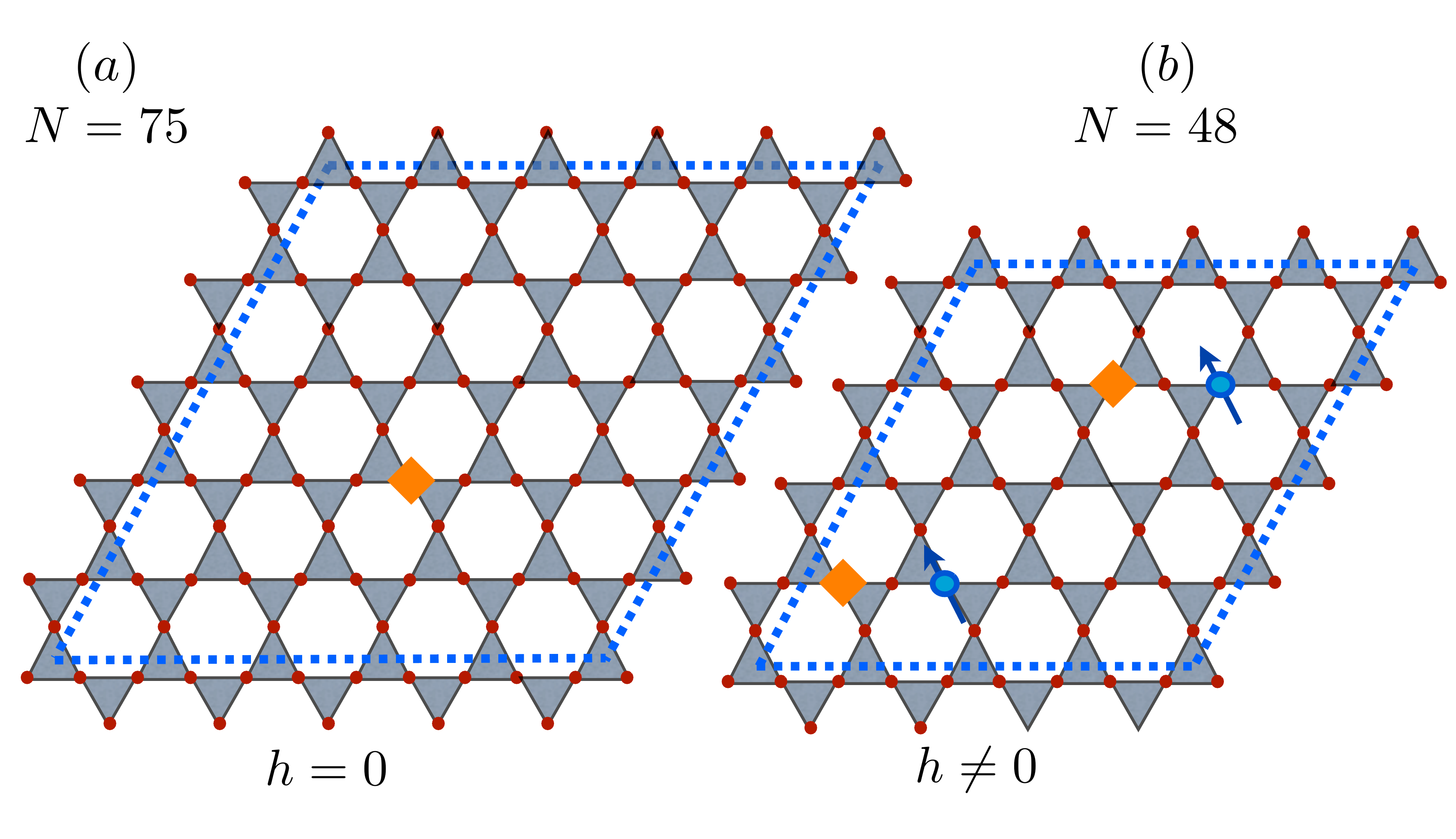}
\end{center}
\caption{Periodic clusters used in this study; (a) The $5\sqrt{3}\times 5\sqrt{3}$ ($N=75$) cluster with a single static impurity.
(b) The $4\sqrt{3}\times4\sqrt{3}$ ($N=48$) cluster doped with $N_{\rm imp}=2$ static impurities and $N_{\rm sp}=2$ mobile spinons
(polarized along the field).}
\label{Fig:clusters}
\end{figure}

{\it Spinon chemical potential --} The density of spinons $n_{\rm sp}=N_{\rm sp}/N$ is tuned by the magnetic field $H$. Indeed, $N_{\rm sp}$ spinons 
polarized along the field gain Zeeman energy $-(h/2)N_{\rm sp}$, where $h=\frac{4}{3}g\mu_B H$, which is balanced by a creation energy term
$(\Delta_{\rm B}/2) N_{\rm sp}$. Note that we rescale here the magnetic field by the same 4/3 factor as the QDM~\cite{schwandt,note-rescale}. Physically, the parameter $\Delta_{\rm B}$ corresponds 
to the energy cost of breaking a singlet 
bond into a triplet made of two polarized {\it static} NN spinons~\cite{normand}. Practically, $(h-\Delta_{\rm B})/2$ plays the role of a chemical potential for the spinons
and the (grand canonical) GS energy becomes,
\begin{equation}
E_{\rm gs} (h,N_{\rm imp})  = {\rm Min}_{/N_{\rm sp}}\{E_{\rm eff} [N_{\rm sp},N_{\rm imp}] -\frac{1}{2}(h-\Delta_{\rm B})N_{\rm sp}\} \, ,
\label{Eq:gs-energy}
\end{equation}
where $E_{\rm eff} [N_{\rm sp},N_{\rm imp}]$ is the GS energy of the QDM depicted in Fig.~\ref{Eq:heff} doped with
$N_{\rm sp}$ {\it mobile} spinons and $N_{\rm imp}$ (random) impurities and where a minimization over $N_{\rm sp}$ has been carried out.
Although, strictly speaking, one expects $\Delta_B=1$ (in units of $J$)~\cite{note-rescale} within the procedure of Ref.~\onlinecite{schwandt} to derive the effective
QDM, its "optimum" value for a given order of the expansion scheme is unclear. 
However, $\Delta_{\rm B}$ can also formally be related to the physical spin gap $\Delta_{\rm S}$ of the {\it pure system}, i.e. the
minimal energy cost to create a triplet ($S=1$) two-spinon state in zero magnetic field (see below).
Lastly, we note that two spinons can form a (triplet) bound state~\footnote{A priori binding of two spinons does not necessarily require asymptotic confinement and can take place in a
deconfined spin liquid phase.}  whenever $E_{\rm eff}[2,0]  + E_{\rm eff}[0,0] -2 E_{\rm eff}[1,0] <0$ (in the pure system).

\section{Single impurity results}
\label{Sec:single}

\subsection{Local dimer modulation}
As a preliminary study, we now start by just inserting a single spinless impurity (no spinon). As stated above, we use the modelisation of the impurity 
in terms of a vacant site as shown in Figs.~\ref{Fig:intro}(b,c). 
In the framework of the generalized QDM, the vacancy suppresses the resonant and diagonal terms of 
Fig.~\ref{Eq:heff} whose loop
contains the vacancy site~\footnote{Other local resonances induced by the impurity itself exist but have negligible effects.}. 

\begin{figure}[h]
\includegraphics[width=0.50\textwidth,clip]{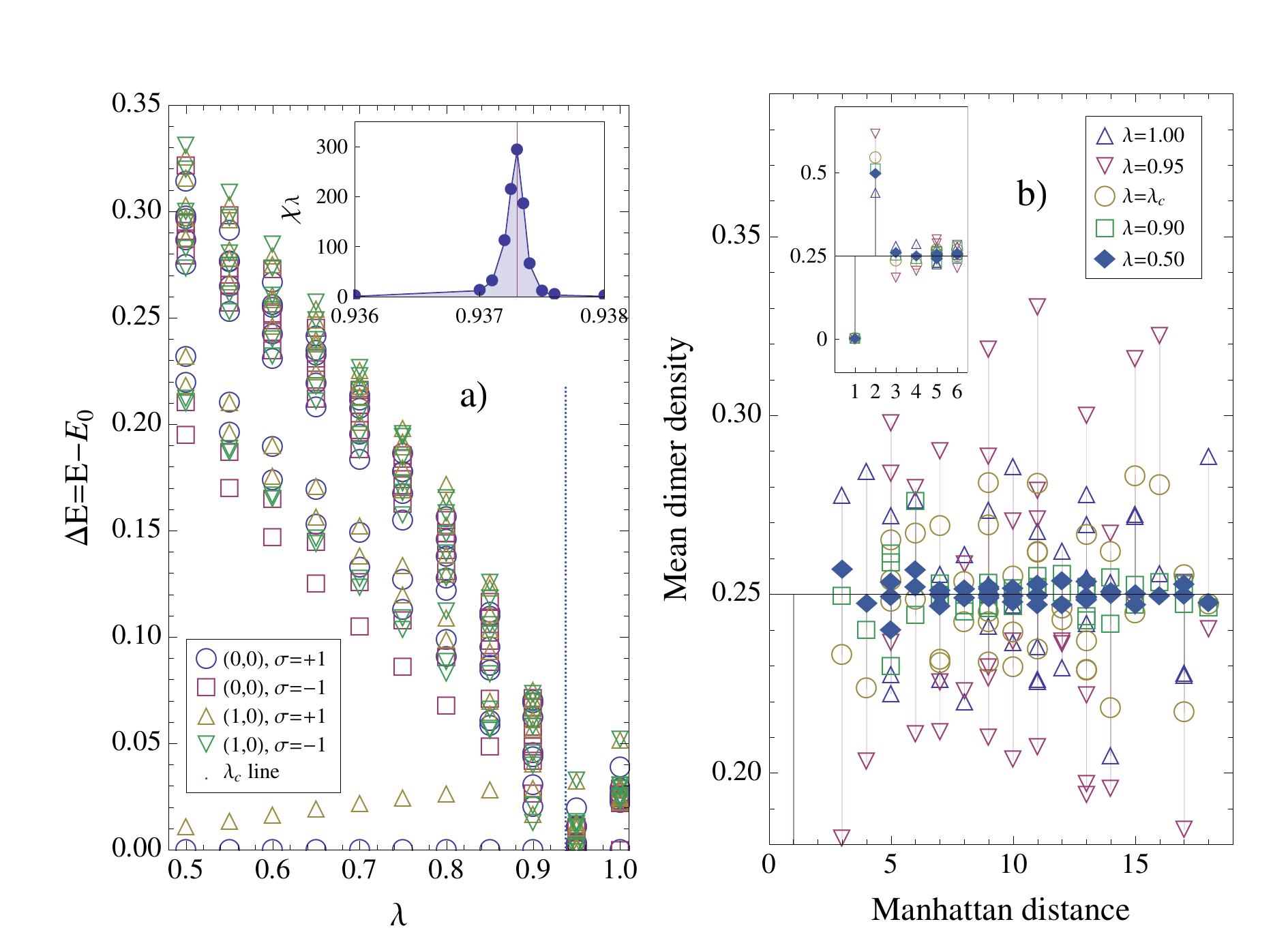}
\caption {(color online)
(a) Spectrum  versus $\lambda$ of the effective QDM on a N=75 site periodic
cluster containing a single impurity and with zero magnetization. The 8 lowest energy levels are
shown in each of the 4 symmetry sectors.  Inset: generalized susceptibility
(see text) showing the location of the phase transition.  For
$\lambda<\lambda_{\rm c}$, the large spectral gap above the quasi-degenerate
topological GS correspond to the {\sl single}-vison gap of the $\mathbb{Z}_2$ liquid.
(b) Average dimer density versus Manhattan distance from the impurity on the
same N=75 cluster. The short distance behavior is separated in the inset for
clarity.  The mean value corresponds to 1/4 dimer per bond. 
}
\label{Fig:ZeroField}
\end{figure}

In the very diluted impurity limit, one does not expect to fundamentally modify the {\sl bulk} properties of the quantum magnet. 
However, a single impurity can e.g. pin VBC order and, in some way, offers a very interesting probe of its existence. 
To test this, we have considered the $5\sqrt{3}\times 5\sqrt{3}$ periodic cluster of Fig.~\ref{Fig:clusters}(a) with $N=75$ sites and
a single impurity (to comply with the hard-core dimer constraints such a cluster has to contain an odd number of 
sites)~\cite{note1}.
The full spectrum of this cluster is shown in Fig.~\ref{Fig:ZeroField}(a) as a function of the parameter $\lambda$. Here we have distinguished between the four topological sectors 
(for such a torus topology)~\footnote{The (0,1), (1,0) and (1,1) sectors are degenerate. See Ref.~\onlinecite{poilblanc}.} and used the reflection symmetry
w.r.t. one of the $C_2$-axis passing through the impurity site ($\sigma=\pm 1$).  A quantum phase transition~\cite{sachdev}  occurs as a function of $\lambda$ for $\lambda=\lambda_{\rm C}\simeq 0.9373$ as evidence from the sharp peak
in the GS generalized susceptibility $\chi_\lambda=\partial^2 E_{\rm GS}/\partial\lambda^2$ shown in the inset~\cite{fidelity}. 
A topological $\mathbb{Z}_2$ liquid (with four-fold degeneracy on a torus) is stable for $\lambda<\lambda_C$.
At $\lambda=\lambda_{\rm C}$ the {\sl single-vortex} (or "vison")
gap~\footnote{This contrasts with the case of the pure system where the spectral
gap corresponds to a {\sl double-vison} excitation. See Ref.~\onlinecite{poilblanc}.} 
corresponding to the first odd ($\sigma=-1$) energy excitation  vanishes. For $\lambda>\lambda_{\rm C}$ the system spontaneously breaks reflection symmetry (the GS is two-fold degenerate 
with $\sigma=\pm 1$ quantum numbers) as expected in the 36-sites VBC.
These results and, in particular, the estimation of the critical value $\lambda_{\rm C}\simeq 0.9373$ are fully consistent with the results obtained on a pure 108-sites cluster~\cite{poilblanc}. 

It is interesting to examine the behavior of the dimer density as a function of the Manhattan distance (see
Fig.~\ref{Fig:intro}(c)) from the impurity, as reported in Fig.~\ref{Fig:ZeroField}(b) for different values of $\lambda$ between
0.5 and 1. One striking feature is the large value of the dimer density on the two bonds facing the impurity (distance $d=2$).
This can be interpreted as a (singlet) {\sl dimer crystallization} as found in prior studies of
the Heisenberg quantum antiferromagnet~\cite{dommange,sindzingre,ioanis}. Our study reveals that this phenomena is very robust and depends
weakly on the supposed phase, liquid ($\lambda<\lambda_{\rm C}$) or crystalline  ($\lambda>\lambda_{\rm C}$), 
of the model. Therefore this feature cannot be used practically as an experimental fingerprint.
In contrast, when considering longer distances, strong differences in the bond modulation occur in the supposed
liquid and VBC phases. For $\lambda<\lambda_{\rm C}$, in the $\mathbb{Z}_2$ dimer liquid, the dimer density becomes very uniform beyond
distance 1. When increasing $\lambda$ the appearance of strong modulations is exactly correlated to the crossing of the critical value
and the entering in the VBC phase. For $\lambda=1$, the relative bond modulation amplitude is of order 20$\%$.

\subsection{Effect of a next-nearest neighbor exchange coupling}

Recent calculations using variational Gutzwiller-projected wavefunctions have shown that a small next-nearest neighbor (NNN) ferromagnetic 
coupling $j_2$ (measured in units of the NN exchange) may favor VBC ordering~\cite{marston-ferro}. A ferromagnetic coupling
should then enhanced the dimer response around impurities, an effect
which can be tested within the QDM framework by considering again the $N=75$ cluster doped with a single impurity (no spinon).
From Ref.~\onlinecite{schwandt}, the new QDM amplitudes $J_\gamma$ and $V_\gamma$, $\gamma=6,8,10$ have simply to be multiplied by 
extra coefficients $(1+A_\gamma)$ which depend on $j_2$.
$A_\gamma$ is given by the ratio of the corresponding QHAF Hamiltonian matrix elements and, from  Eq.~(12) of Ref.~\onlinecite{schwandt}
giving the general form of these matrix elements, one gets $A_\gamma=j_2\sum_{(i,j)\in\gamma,\hbox{\tiny NNN}}\epsilon_{i,j}/(\gamma/2+\sum_{(i,j)\in\gamma,\hbox{\tiny NN}}\epsilon_{i,j}$)
where each loop topology has now to be distinguished and $\epsilon_{i,j}=-1$ ($+1$) if the sites $i$ and $j$ on the loop $\gamma$ 
are separated by an odd (even) distance. Note that for $\gamma=12$ (star resonance), 
the NN QHAF matrix element $\gamma/2+\sum_{(i,j)\in\gamma,\hbox{\tiny NN}}\epsilon_{i,j}$ vanishes so that the leading contributions to $J_{12}$ and $V_{12}$ are directly
proportional to $j_2$.
Using the general expansion scheme given by Eq.~(41) of Ref.~\onlinecite{schwandt} it is straightforward to obtain the
leading contribution (at order $\alpha^{10}$ and $\alpha^{20}$ respectively), $J_{12}=-\frac{3}{8}j_2$ and $V_{12}=-\frac{3}{256}j_2$.

\begin{figure}
\includegraphics[width=1.\columnwidth,clip]{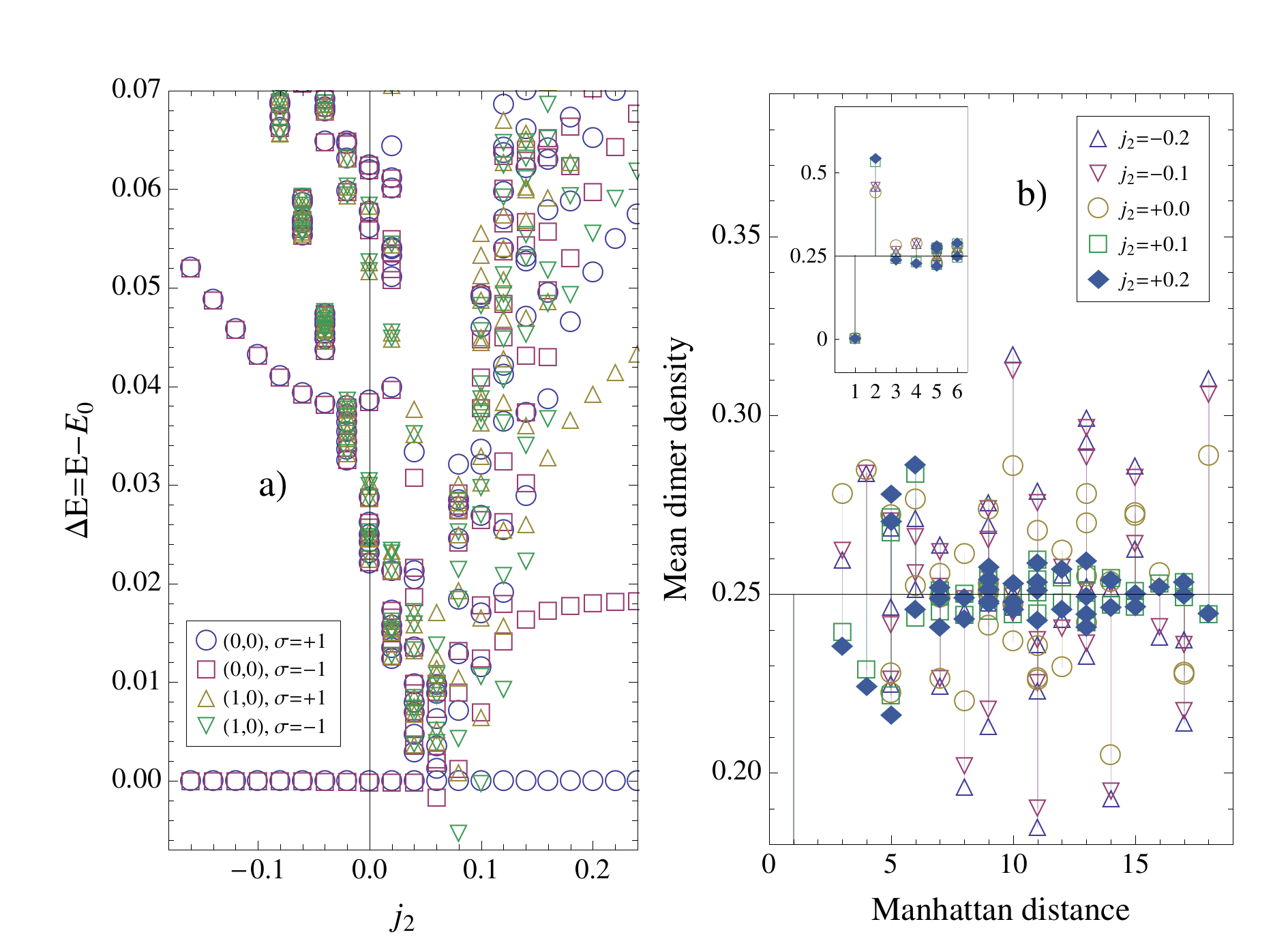}
\caption {(color online)
(a) Spectrum  versus $j_2$ (NNN coupling in units of the NN coupling) of the $\lambda=1$ effective QDM on a N=75 site periodic
cluster containing a single impurity and with zero magnetization. The lowest energy levels are
shown in each of the 4 symmetry sectors.  
(b) Average dimer density versus Manhattan distance from the impurity on the
same N=75 cluster. The short distance behavior is separated in the inset for
clarity.  The mean value corresponds to 1/4 dimer per bond. 
\label{Fig:j2}
}
\end{figure}

The excitation spectrum for $\lambda=1$ shown as a function of $j_2$ in Fig.~\ref{Fig:j2}(a) reveals two gapped phases separated by
a narrow gapless region located around $j_2^c\sim 0.05$ (which might become a quantum critical point in the thermodynamic limit).  
The $j2<j_2^c$ region (including $j_2=0$) corresponds to the above pinned VBC as confirmed by the GS quasi-degeneracy and the 
enhanced VBC patterns shown in Fig.~\ref{Fig:j2}(b) for weak {\it ferromagnetic} $j_2$. In contrast, for AFM NNN exchange $j_2>j_2^c$,
only a small dimerization is observed in  Fig.~\ref{Fig:j2}(b), except at short distances from the impurity. 
This might be the signature of a dimer liquid or a very weakly pinned VBC
and more work (on pure systems) is needed to resolve this case.

\subsection{Does a spinon bind to an impurity ? }

\begin{figure}
\includegraphics[width=1.0\columnwidth,angle=0]{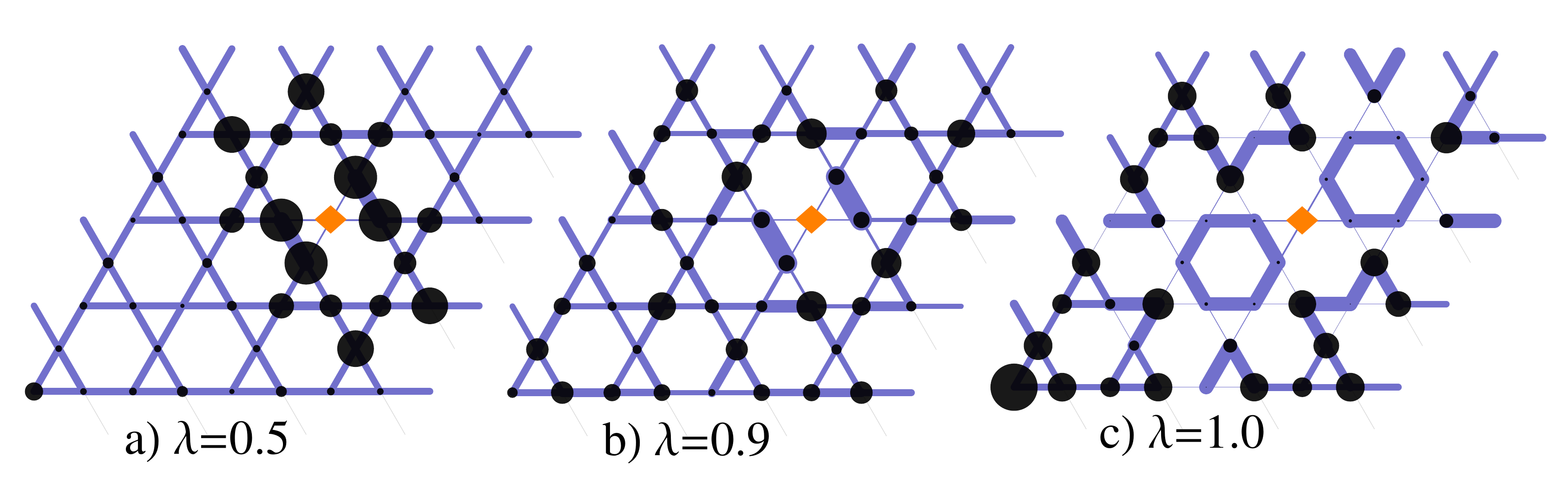}
\caption {(color online)
Real-space maps of the the  dimer density (segments)
and site magnetization (circles) computed on a periodic $N=48$-site cluster doped with a single impurity
and a mobile spinon, for different values of $\lambda$ (as shown on the plots).
The radii (width) of the circles (segments) are directly proportional to the values of the corresponding observables. }
\label{Fig:maps}
\end{figure}

Notoriously, spinons in dimer liquids (VBC) are known to be deconfined (confined). However, confinement/deconfinement is an asymptotic (e.g. long-distance) property of the system
and unexpected behaviors can appear at shorter distance (comparable e.g. to the typical impurity separation).
Indeed, a spinon could bind to an impurity (acting here as a "static spinon") even in a dimer liquid or, reversely, a spinon could spread 
over a very large area
(e.g. beyond the cluster sizes available numerically) even in a VBC.
To investigate such issues, we now consider a $4\sqrt{3}\times4\sqrt{3}$ 
(48-sites) cluster~\cite{note1} ($j_2=0$) doped with a static vacancy and a mobile spinon. 
The maps of both the (bond) dimer density and the site magnetization
are shown in Fig.~\ref{Fig:maps} for three distinct regimes~: (i) $\lambda=0.5<\lambda_{\rm C}$, deep in the dimer liquid
phase;  (ii) $\lambda=0.9<\lambda_{\rm C}$ and (iii) $\lambda=1> \lambda_{\rm C}$, in the dimer liquid and VBC 
phases respectively, in the vicinity of $\lambda_C$.
The dimer density maps in this lightly polarized system closely resemble the previous zero field data
of Fig.~\ref{Fig:ZeroField}(b)~: the dimer liquid phase at $\lambda=0.5$ exhibits a very homogeneous dimer density 
apart from the two "crystallized" bonds next to the impurity while strong dimer patterns are clearly visible in the VBC phase. 
Note that  the dimer pattern gets more pronounced in the dimer liquid phase 
when approaching the dimer liquid-VBC transition, as e.g. for $\lambda=0.9$.
The local magnetization maps of Figs.~\ref{Fig:maps}(b) $\&$ (c) are entirely correlated with the dimer density 
maps: for $\lambda=0.9$ (liquid)
the magnetization is slightly depressed on the four bonds NN to the impurity and very uniformly distributed beyond.
For $\lambda=1$ (VBC) a strong modulation in the local magnetization appears.
In other words, the magnitudes of the dimer and of the (magnetic field induced) spin density modulations seem to scale with each other. 
The behavior in the dimer liquid phase is fully consistent with the expected {\it spinon deconfinement} mechanism
found by LED of small clusters of the Kagome QHAF~\cite{dommange,deconfine}. 
In contrast, the VBC sample does not seem to show {\it at the length scale of the cluster} 
the (opposite) confinement behavior found e.g. in the VBC phase of the QHAF on the
checkerboard lattice~\cite{deconfine}. 

Although probably not directly relevant to experiments, the data of Fig.~\ref{Fig:maps}(a), deep in the dimer liquid phase, are of particular theoretical interest~: here, although 
the dimer pattern is very uniform (as expected in a dimer liquid phase) and no confining potential
is expected, the spinon still forms a tight bound state with the impurity. We believe however that this behavior is not generic and depends strongly on the
details of the spinon Hamiltonian~: here the bound state is stabilized by an increase of  spinon kinetic energy in the vicinity of the impurity. 
Whether such bound state can appear spontaneously without magnetic field depends on the value of the spinon chemical potential 
i.e. on $\Delta_B$. If this is the case, the impurity-doped system becomes gapless for magnetic excitations (see below).

\section{Finite impurity concentration}
\label{Sec:finite}

We now consider the case of $N_{\rm imp}=2$ impurities doped into the same cluster of $N=48$ sites, attempting to (crudely) mimic a finite
density of impurities $n_{\rm imp}=N_{\rm imp}/N$ of order $4\%$. For simplicity, we fix the relative position between the two impurities
at a "typical" distance, as shown in Fig.~\ref{Fig:clusters}(b). We shall also consider the same values of the parameter $\lambda$ as above in the investigation of a single impurity, 
namely $\lambda=0.5$, $0.9$ and $1$, for comparison. Hereafter, we also set $j_2=0$.

\subsection{Enhanced dimerization}

\begin{figure}
\includegraphics[width=1.0\columnwidth,angle=0]{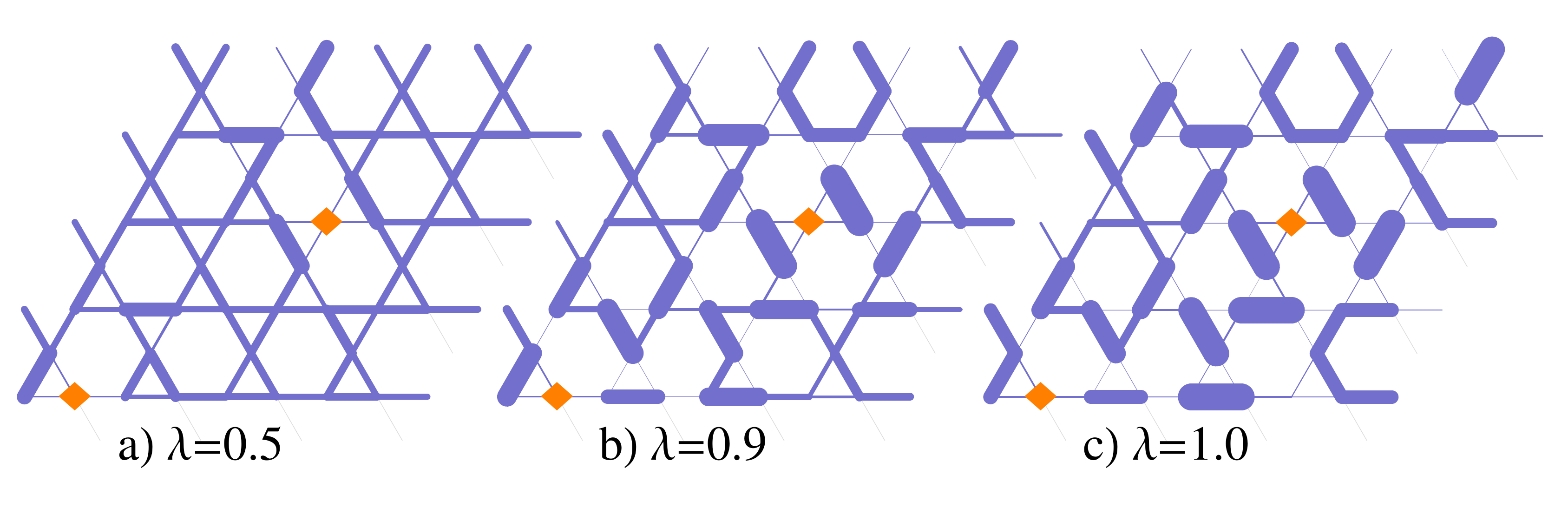}
\caption {(color online)
Real-space maps of the dimer density (segments)
on a periodic $N=48$-site cluster doped with two impurities. Same conventions as Fig.~\ref{Fig:maps}.
}
\label{Fig:maps2}
\end{figure}

Let us first assume zero total magnetization e.g. $N_{\rm sp}=0$. The corresponding data are shown in Figs.~\ref{Fig:maps2}(a-c).
We clearly observe a sizable increase of the average dimerization compared to $N_{\rm imp}=1$. 
This is particularly clear for $\lambda=0.9$, a parameter corresponding, in the pure system,
to the dimer liquid phase close to the phase transition at $\lambda=\lambda_C$.  In fact, with two doped impurities, the new dimerization pattern in Fig.~\ref{Fig:maps2}(b) 
resemble very closely the one at $\lambda=1$ in Fig.~\ref{Fig:maps2}(c). 
It is interesting to also note that the local dimerization patterns around impurities can be different. Indeed, e.g. in Figs.~\ref{Fig:maps2}(b) $\&$ (c), while one impurity
is surrounded by two (quasi-isolated) strong dimer bonds, the second impurity stabilizes two neighboring symmetric resonating hexagons. 
All these findings are in fact consistent with the proposed picture of the Valence Bond Glass~\cite{singh-glass}.

\subsection{Spin gap}

\begin{figure}
\includegraphics[width=0.45\textwidth,clip]{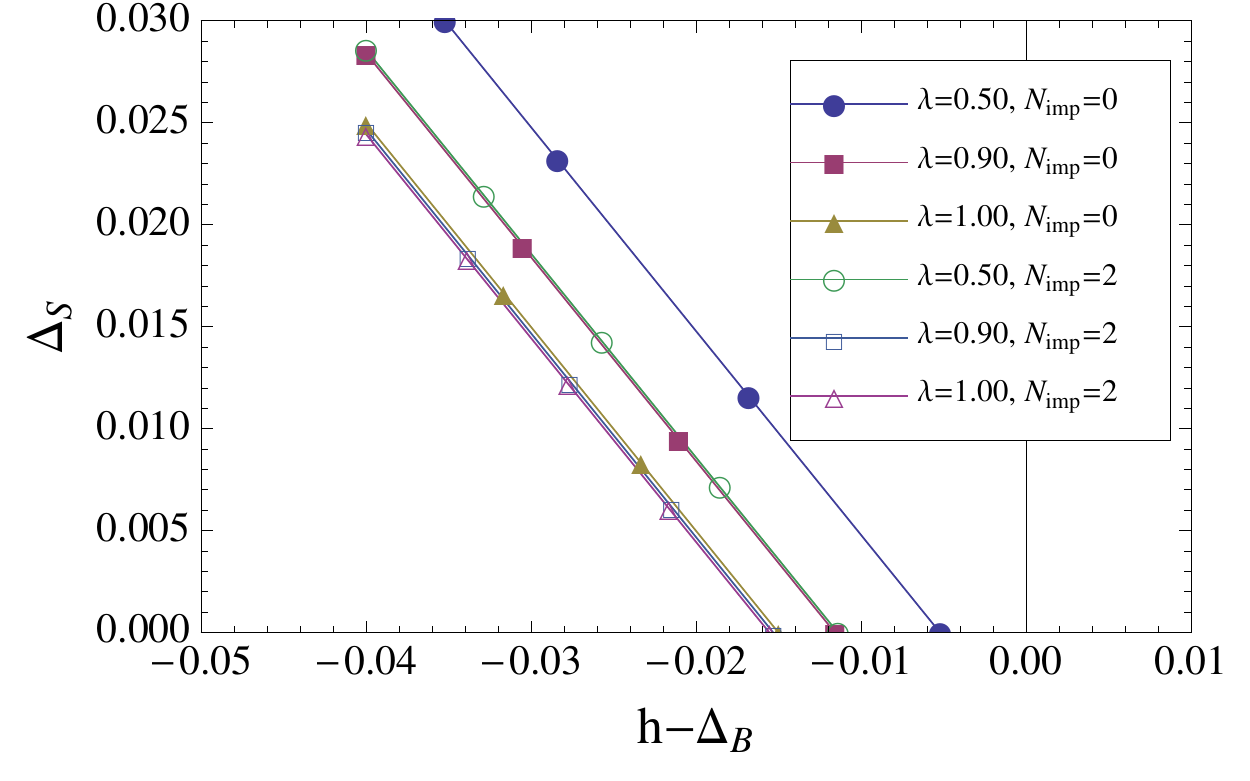}
\caption {(color online)
Spin gap of the pure (filled symbols) and two-impurity (empty symbols) doped 48-site cluster vs the parameter $\Delta_B$. }
\label{Fig:SpinGap}
\end{figure}

We now considerer the possibility that the ground state carries a small magnetization proportional to the
(doped) spinon density $m=\frac{1}{2}n_{\rm sp}$. If a spin gap exists at $h=0$, we expect that 
a large enough magnetic field will eventually suppress the gap and induce a small magnetization. 
In our picture, the spin gap is the minimal energy to create a pair of
mobile spinons. From  Eq.~(\ref{Eq:gs-energy}), the spin gap at low field $h$ is therefore given by: 
\begin{eqnarray}
\Delta_{\rm S}(h,n_{\rm imp})&=&E_{\rm eff}[2,N_{\rm imp}]  - E_{\rm eff}[0,N_{\rm imp}]  + \Delta_{\rm B} -h\nonumber\\
&=&  \Delta_{\rm S}(0,n_{\rm imp})-h \, ,
\label{Eq:spin-gap}
\end{eqnarray}
where $\Delta_{\rm B}$ can be taken as a phenomenological parameter (considering the fact that our spinon hamiltonian is only a minimal model
and does not include the potentially important higher-order longer-range hoppings).
As in the conventional picture, the spin gap (if any) vanishes linearly at a critical field. 
We have numerically computed the spin gap in the same $N=48$ cluster doped with two impurities (same $\lambda$'s also as above) i.e. for 
an impurity concentration $n_{\rm imp}=N_{\rm imp}/N\simeq 4\%$, and compare it to the equivalent data in the pure 
system ($n_{\rm imp}=0$) in Fig.~\ref{Fig:SpinGap}. 
Interestingly, we find that doped impurities lead to a {\it reduction} of the spin gap in the dimer liquid phase e.g. for $\lambda=0.5$. In the VBC phase
the effect is much less pronounced. We believe that the existence of an impurity-spinon bound state at $\lambda=0.5$ (see above)
is responsible for this trend: if spinons get bound to available impurities (see below)
the energy cost to create two of them is reduced. 

\subsection{Spin density modulations}

\begin{figure}
\includegraphics[width=1.0\columnwidth,angle=0]{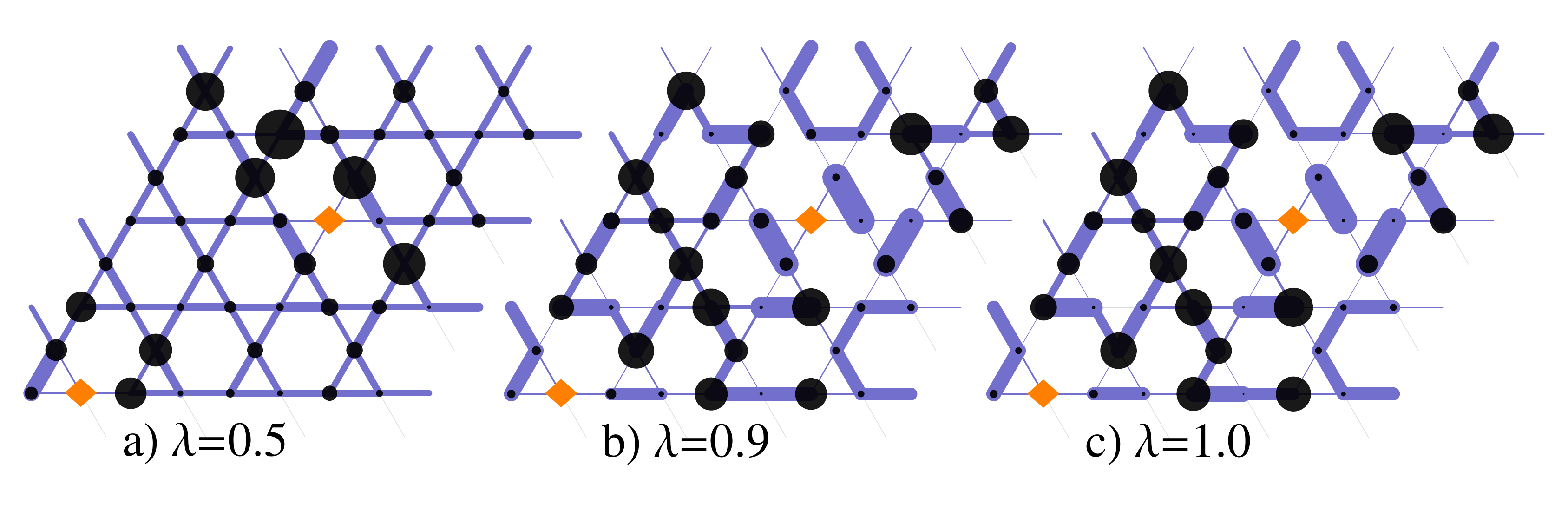}
\caption {(color online)
Real-space maps of the  dimer density (segments)
and site magnetization (circles) computed on a periodic $N=48$-site cluster doped with two impurities
and two mobile spinons. Same conventions as Fig.~\ref{Fig:maps}.
}
\label{Fig:maps3}
\end{figure}

We have also examined the spinon density map (equivalent to the local magnetization map) in 
the two spinon-two impurity doped $N=48$ cluster, again for the same values $\lambda$ as above.
The features shown in Figs.~\ref{Fig:maps3}(a-c) are consistent with our previous results for a single impurity. 
In the $\lambda=0.5$ dimer liquid, a spinon clearly forms a bound state around each 
impurirty. In contrast, in Figs.~\ref{Fig:maps3}(b) $\&$ (c) which we might view as some local region of a
Valence Bond Glass~\cite{singh-glass},
spinons seem to be repelled (at short distance) from the impurities. 
At further distances, strong spin density 
modulations are induced by a strong (average) dimerization (which reversely is weakly affected by the small 
magnetization of the sample).

\section{Application to NMR experiments}
\label{Sec:nmr}

\subsection{General considerations}

\begin{figure}[htb]
\includegraphics[width=0.95\columnwidth,clip]{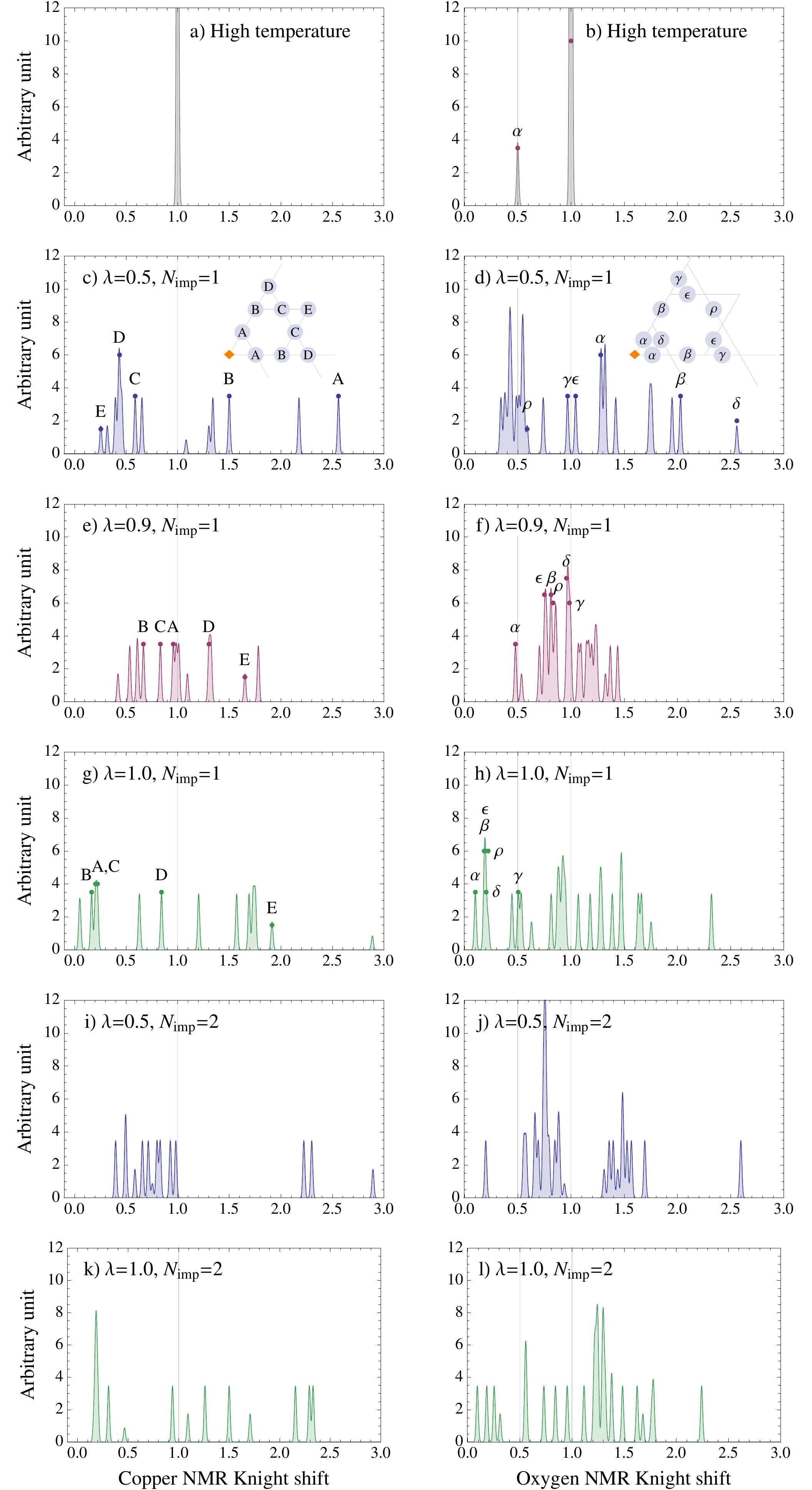}
\caption {(color online)
Copper (left) and Oxygen (right) NMR spectra computed from the magnetization maps of Fig.~\ref{Fig:maps}(a-c) obtained
for a single impurity on $48$ sites (effective concentration of $\sim 2\%$) and of Fig.~\ref{Fig:maps3}(a) and (c) obtained
for two impurities on $48$ sites (effective concentration of $\sim 4\%$).
The Knight shift variables are normalized 
w.r.t. the values corresponding to an ideal uniform distribution of the spin density and defining the reference NMR spectra (a,b) 
labeled as "high-temperature".  Zero-temperature spectra for $\sim 2\%$ impurity concentration 
$\lambda=0.5$ (c,d), $\lambda=0.9$ (c,d) and
$\lambda=1$ (g,h); Same for $\sim 4\%$ impurity concentration 
$\lambda=0.5$ (i,j) and
$\lambda=1$ (k,l).
For a single impurity, the impurity peaks are labelled according to the positions of the resonating
nucleus shown in the insets (see also Fig.~\ref{Fig:intro}(c)). 
\label{Fig:nmr}
}
\end{figure}

The different magnetization maps in the liquid and VBC phases are expected to lead to
different characteristic NMR (theoretical) spectra which can be confronted to experiments
in Herbertsmithite~\cite{nmr}. The theoretical Copper NMR spectra is defined by the histogram of the local site magnetizations.
Since Oxygen is located between two coppers, the effective local magnetization is obtained by summing up the
contributions from both sites (named here as {\sl bond} magnetization). Both site and bond magnetization histograms have been
computed for $\lambda=0.5$, $\lambda=0.9$ and $\lambda=1$ using the spinon density maps of
Figs.~\ref{Fig:maps}(a-c) (one impurity on $48$ sites) and Figs.~\ref{Fig:maps3}(a) $\&$ (c) (two impurities on $48$ sites)
providing the theoretical Copper and Oxygen NMR low-temperature spectra shown in Figs.\ref{Fig:nmr}. 
Note that, for convenience, we work here at fixed magnetization~\footnote{Instead, imposing the magnetic field 
would require the knowledge of the field (and temperature) dependence of the macroscopic magnetization.} 
corresponding to one spinon per impurity i.e. $m=\frac{1}{2}n_{\rm imp}$.
This is a natural choice at very small temperatures if spinons bind to impurities or if the spin gap is very small. 
In any case, this corresponds physically to a magnetic field $h>\Delta_S(h=0)$.
For convenience, the Copper and Oxygen Knight shifts (x-axis) have been normalized w.r.t. the
bulk values corresponding to a uniform spatial distribution of the magnetization (mimicking high temperatures). 
Note however that, even for such a uniform distribution of the Copper spin density, 
the Oxygen nucleus located on the bonds connected to the impurity site
see only one Copper atom instead of two, yielding a small "satellite" peak 
in the spectrum of Fig.~\ref{Fig:nmr}(b) at half the bulk Oxygen NMR shift. 

Comparing quantitatively high and low-temperature (in fact here $T=0$) spectra 
obtained at a low {\it fixed magnetic 
field} would in principle require the knowledge of the
temperature dependent magnetic susceptibility $\chi(T)$ which is beyond our numerical capability. 
However, a number of robust features of these theoretical NMR spectra can be linked to physical behaviors of the spinons around impurities.
E.g. when spinons spontaneously bind to impurities (as it is the case in the $\lambda=0.5$ dimer liquid phase within our model),
one expects low-energy magnetic excitations, possibly at arbitrary small energy scale so that the broad theoretical spectra 
(at fixed $m$) of Fig.~\ref{Fig:nmr}(c,d,i,j) should provide a good sketch of {\it low-field} NMR spectra. 
On the other hand, Figs.~\ref{Fig:nmr}(e,f) correspond to the opposite case where spinons are strongly delocalized around 
impurities leading to narrower structures in the spectra (in the same units). However, if the system retains a finite spin gap
(as expected in this type of dimer liquid), for a {\it small magnetic field}, the average position of the spectrum should be strongly 
pushed downwards w.r.t. its high temperature reference values.

\subsection{Herberthsmithite described as a Valence Bond Glass} 

Let us now discuss our results in the light of the recently obtained $^{17}$O NMR experimental spectra
on the Kagome QHAF Herberthsmithite material which contains an intrinsic small Zinc substitution in the Copper Kagome planes~\cite{nmr}. 
Although the experimental NMR spectra are strongly broaden by quadrupolar effects typical of powders, two separate structures can clearly be followed from high to low temperatures. 
The average shift of the "satellite" peak is roughly half of the main one so that this peak has been interpreted as resulting from the resonance of the four Oxygen nucleus (labelled as "$\alpha$" in Fig.~\ref{Fig:nmr}) next to an impurity site. 
Our results show that dimerization is enhanced by impurity doping so that the proposed Valence Bond Glass~\cite{singh-glass} should be stabilized in an enlarged interval of $\lambda$ compared to the region of stability of the VBC in the pure model. 

It has been recently argued that, confined spinons in the Valence Bond Glass can form a
random-singlet phase with a wide distribution of spin gaps down to zero energy, ensuring the presence of fluctuating 
spins at arbitrarily low temperatures as indeed observed experimentally.  
The low-temperature low-field magnetic response of a Valence Bond Glass is therefore dominated by 
{\it isolated} impurities in rare regions of the sample, each of them carrying an almost free spinon
loosely bound by a (shallow) VBC confining potential~\footnote{In that case, the bound state size is relatively large so that
the $48$ site cluster with a single impurity
is still too small to observe real confinement.}. Within this scenario the $T,h\rightarrow 0$ NMR spectrum
is qualitatively given by the dominant magnetic response of the above regions. If this physics is at play in Herberthsmithite, 
Fig.~\ref{Fig:nmr}(h) should therefore give a crude qualitative sketch of the $T\rightarrow 0$ Oxygen NMR spectrum. 
This spectrum vaguely shows two broad structures (as experiments)
but the assignment of the small shift structure to the $\alpha$ Oxygen is not so clear as shown in Fig.~\ref{Fig:nmr}(h).
Note that the NMR spectrum would become more complicated when raising temperature
since more and more regions of the sample will contribute (as e.g. Fig.~\ref{Fig:maps3}(c) giving Fig.~\ref{Fig:nmr}(l)).

\section{Discussions and conclusions}
\label{Sec:conclusions}

Theoretically, our approach is certainly based on an approximate framework relying
on i) the truncation within the NN VB basis, ii) an overlap expansion up to a finite order and iii) a "minimal" spinon effective
hamiltonian. However, we believe it still captures a number of important features and
unambiguously shows that even a small (experimentally unavoidable) impurity doping should have important experimental consequences. Fig.~\ref{Fig:maps3}(c) probably gives a fairly reliable picture of what is to be expected in a small area 
of the experimental system at finite magnetization. Even an impurity 
concentration as small as $4\%$ leads to an important increase of the average dimerization and to
a fairly inhomogeneous spin density distribution. This property should be robust 
in a more elaborate treatment of the NN QHAF, like for example when using a refined spinon Hamiltonian including longer-range hoppings. 

It has been recently argued that, confined spinons in the VBC can form a
random-singlet phase with a wide distribution of spin gaps~\cite{singh-glass}, ensuring the presence of fluctuating spins at arbitrarily
low temperatures as indeed observed experimentally.  
Interestingly, our data on small clusters show that the VBC pattern is not strongly confining, probably due to the weakness of the
VBC amplitude compared to the spinon kinetic energy. On the scale of the typical impurity separation,
spinons tend to delocalize around impurities. 
However, different types of dimerization patterns get pinned around different impurities.
In this sense, our findings have some similarity with the picture of the Valence Bond Glass~\cite{singh-glass}. 
Not surprisingly, we have not found any appreciable reduction of the spin gap (compared to the pure case) 
for the (two-) impurity configuration we have chosen (except when spinon-impurity bound states form at $\lambda=0.5$)~:
we believe our cluster is still too small to realize (rare) configurations  with isolated impurities
that could exhibit (almost) free spin within some (large) confining length. This issue certainly needs further work. 

On the experimental side, a more realistic description of Herbertsmithite may require
to go beyond the simple NN QHAF. It is possible that longer range exchange interactions destabilize the VBC ground state in favor of the $\mathbb{Z}_2$ dimer liquid or some other candidate GS. 
It might also happen that the VBC phase is only stable at temperatures lower than those experimentally reached. 
Finally, we note that, experimentally, a large enough DM anisotropy~\cite{dm} could fill out the spin gap of the  $\mathbb{Z}_2$ dimer liquid or VBC phases.

{\it Acknowledgements} -- We thank M.~Mambrini, G.~Misguich and R.~Singh for valuable
discussions. We are also indebted to Fabrice Bert and Philippe Mendels for help in the understanding
of their NMR data.  D.P. acknowledges IDRIS (Orsay, France) for allocation of
CPU time on the NEC supercomputer and A.R. acknowledges CIMENT (Grenoble,
France) for access time to the local computation cluster.

 

\begin{thebibliography}{99}
\bibitem{note1} The $48$ site and $75$ site clusters used here possess all the
point group symmetries of the (infinite) kagome lattice. Although they do not
accommodate the VBC 36-sites cell, they still capture correctly the local
physics around impurities. The largest Hilbert spaces are obtained for a single
impurity in $75$ sites ($4\, 194\, 304$ symmetrized states) and for 2 spinons
and 2 impurities in $48$ sites ($8\, 511\, 488$ states).
\bibitem{note-rescale}  We extend the rescaling of the QHAF hamiltonian of Ref.~\onlinecite{schwandt} to sectors with finite $S_Z=N_{\rm sp}/2$ as
$\hat{\cal H}_H\rightarrow (4/3)\hat{\cal H}_H+J(N-N_{\rm sp})/2$.
\end{thebibliography}
\end{document}